\documentclass[12pt]{article}
\usepackage{amssymb,amsmath}
\usepackage{graphicx}

\begin{document}
\title{Coupled-channel analysis of collisional effects on HFS transitions in
antiprotonic helium atoms}
\author{G.Ya. Korenman and S.N. Yudin \\
\itshape Institute of Nuclear Physics, Moscow State University,
Moscow, Russia \\ e-mail: korenman@anna19.sinp.msu.ru}
\date{}
\maketitle

\begin{abstract}
 \noindent
Collisions of metastable antiprotonic helium with atoms of medium
induce transitions between hyperfine structure sublevels as well
as shifts and broadenings of the microwave M1 spectral lines. We
consider these phenomena in the framework of a simple model with
scalar and tensor interactions between
$(\bar{p}\mathrm{He}^+)_{nL}$ and $\mathrm{He}$ atoms. $S$-matrix
is obtained by solving coupled-channels equations involving 4 HFS
sublevels $(F=L\pm 1/2,\,J=F\pm 1/2)$ of the $nL$ level and
relative angular momenta up to $l=5$ at the kinetic energy
$E\lesssim 25$ K. The calculated spin-flip cross sections are less
than elastic ones by four orders of value in the cases
 $\Delta F=\Delta J=\pm 1$ and $\Delta F=0,\,\Delta J=\pm 1$,
and by seven orders of value in the case $\Delta F=\pm 1,\, \Delta J=0$
or $\Delta J=\pm 2$. The considered cross sections reveal a resonance
behaviour at very low energy ($E\sim 1\div 4$ K depending on the model
parameters). At the density $N=3\times 10^{20}$ cm$^{-3}$ and $T=6$ K
we obtain the relaxation time $\tau(FJ\rightarrow F'J')\geq 160$ ns,
the frequency shift $\Delta\nu\simeq 80$ kHz and the frequency
broadening  $\gamma \lesssim 5.9$ MHz for M1 spectral lines  of the
favored transitions ($\Delta F=\pm 1,\, \Delta J=\pm 1$). The results
are compatible with the recent experimental data obtained by a
laser-microwave-laser resonance method. With the temperature rising up
to 25 K the rate of relaxation $\lambda=1/\tau$ as well as shift and
broadening of the M1 microwave lines are lowered by a factor $1.5 \div
2 $ reflecting the displacement of a mean kinetic energy from the
region of resonance scattering.
\end{abstract}

              \section{Introduction}
Up to 1991 it was generally accepted that antiprotons after
entering a matter would very quickly annihilate via strong
interaction with nuclei of matter. However in 1991 at KEK it was
discovered \cite{1} that antiprotons could really survive in the
midst of the ordinary helium atoms for the time intervals at least
of the same order of value as  the lifetime of the muon. This
phenomenon is explained by the formation of the specific
metastable states, for which radiative and Auger transitions are
strongly suppressed. Similar states for hadronic helium atoms were
predicted a long time ago \cite{2}. Internal structure of the
$(\bar{p}\mathrm{He}^+)^*$ metastable states can be considered,
roughly, as electron in the ground state and antiproton in the
$nL$ -state with $ n\approx n_0=\sqrt{M^*/m_e}\approx 38$, where
$M^*$  is a reduced mass of $(\bar{p}-\mathrm{He}^+)$ system. The
angular momentum $L$ should be sufficiently large to form a
circular or nearly-circular state, i.e. $L\approx n-1$.

If antiproton has "sat down" on a such orbit, it will experience
only a delayed annihilation. Therefore the sequence of the
phenomena after entering of the antiprotons in matter seems to be
as follows. When the antiproton will be proved to be in the helium
target it quickly loses its energy through the ionization. When
the kinetic energy falls below the ionization energy of helium
atom ($ I=24.6$ eV), the antiproton knocks out one of the
electrons in a He atom by the Coulomb interaction and occupies the
orbit that is close to the orbit of the knocked electron. The
formed $(\bar{p}\mathrm{He}^+)^*$ atom should have an initial
kinetic energy about $5$ eV as a recoil energy and after a time
shorter than a nanosecond it reaches the thermal equilibrium with
the medium.

 After thermalization the antiprotonic atom will be in one of the
states accessible for it and will proceed to collide with the
neutral atoms of helium. These collisions will change the initial
populations of the states of the antiproton atoms and also other
their properties, such as rates of transitions, shifts and
broadenings of spectral lines. These changes were identified by
the method of laser resonance spectroscopy so it is opened a new
layer of interesting physics (see \cite{3} and the references
therein). In particular, the density shifts and broadenings of E1
spectral lines $(nL\rightarrow n'L'=L\pm 1)$ were observed for
laser-induced transitions. The model theoretical analysis \cite{4}
shows that qualitative peculiarities of the data are related to
quantum effects at very low temperature and peculiar features of
$(\bar{p}\mathrm{He}^+)-\mathrm{He}$ interaction. More
sophisticated calculations with \emph{ab initio} potential surface
\cite{5} give a quantitative agreement with the experimental data
for a lot of E1 transitions.

Recently, the first data on hyperfine structure of the
$(n,L)=(37,35)$ state of the $(\bar{p}\mathrm{He}^+)$ were
obtained by a laser-microwave-laser resonance method \cite{6}. The
central frequencies of microwave M1-transitions,
$\nu_{HF}^+(F=L-1/2,J=L \rightarrow L+1/2,L+1)=12.89596(34)$ GHz
and $\nu_{HF}^-(L-1/2,L-1 \rightarrow L+1/2,L)=12.92467(29)$ GHz,
are in excellent agreement ($\lesssim 30$ ppm) with the recent
calculations for the isolated $(\bar{p}\mathrm{He}^+)$ system
\cite{7}. The results suppose that the density shifts of the M1
spectral lines at the experimental conditions are very small and
do not exceed the experimental accuracy ($\sim 300$ kHz), in
contrast with the $E1$ transitions. However the width of the lines
($\gamma\sim 5.3\pm 0.7$ MHz) leaves room for a collisional
broadening. One more consequence of the data is that the mean time
of the collisional relaxation $\tau_c(F=L-1/2\rightarrow
F'=L+1/2)\gtrsim 140$ ns, the latter number being the observation
time window.

As far as we know there is no published theoretical papers on
collisional effects on HFS states of the $(\bar{p}\mathrm{He}^+)$
system, except for our short paper \cite{8}.
 We consider the problems in the frame of a simple model of
$(\bar{p}\mathrm{He}^+) - \mathrm{He}$ potential \cite{4} extended
to the tensor interaction. In this paper we investigate the
complex of the mentioned questions by the coupled channel method.

       \section{Formalism and Model}
\begin{figure}[thb]
      \centering
\vspace*{-5mm}
\includegraphics[width=0.5\textwidth]{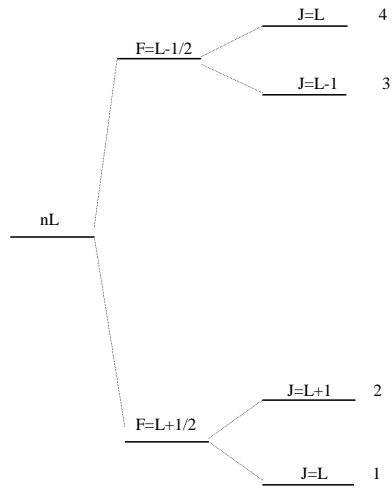}
\vspace*{-20mm}
 \caption{Hyperfine splitting of the  $(\bar{p}\mathrm{He}^+)_{nL}$ level.
 The numbers on the right side from the sublevels are given for the short
references to the states with definite $F$ and $J$. }
 \label{fig:HFS}
\end{figure}

The method of coupled channels consists in the solution of the
finite system of coupled differential equations, which is obtained
from the total Schroedinger equation by developing a total wave
function into a restricted set of the basis channel states. The
number of the equation in this system is defined by the number of
different states of the antiprotonic atom included into
consideration and by the number of different partial waves of
relative motion of colliding atoms at the given kinetic energy.
that is necessary to get the reasonable cross sections. We take
into account only four states arising as a result of spitting of
$nL$ level, due to the interaction of the orbit L with spins of
electron and antiproton. Four levels of the antiprotonic atom that
were included into coupled channel system are shown on Fig.
\ref{fig:HFS}. The wave functions of the states can be written in
a good approximation as a vector coupling of corresponding orbital
and spin angular momenta,
 \begin{equation} \label{1}
|nL,s_e(F)s_{\bar{p}}:JM\rangle =\left( \left(\Phi_{nL}(\xi)\otimes
|s_e\rangle \right)_F \otimes |s_{\bar{p}}\rangle \right)_{JM},
\end{equation}
where $\Phi_{nL}(\xi)$ is a space wave function of the
$(\bar{p}\mathrm{He}^+)$ in the state with a principal quantum
number $n$  and orbital quantum number $L$,  $\xi$ is a set of
inner coordinates of the system, $|s_e\rangle$ and
$|s_{\bar{p}}\rangle$ are spin functions of electron and
antiproton,  $F=L\pm1/2$ and $J=F\pm 1/2$ are intermediate and
total angular momenta of the split substates
($\mathbf{F}=\mathbf{L}+\mathbf{s}_e$,
$\mathbf{J}=\mathbf{F}+\mathbf{s}_{\bar{p}}$).

Further, it is necessary to fix the set of orbital momenta $l$ of
the relative motion of the antiprotonic and helium atoms.  At the
relative kinetic energy $E\leq 25$ K the wave number $k< 0.8$
a.u., and, with an effective radius of the model interaction (see
below), we estimate $l_{\max}=5$. We check also by the
calculations with coupled channel equations that the contribution
of the relative angular momenta $l>5$  into the cross sections
under consideration are negligible.

Now the channels will be defined by the quantum numbers of antiprotonic
states, orbital angular momentum $l$ and total angular momentum $j\,m$,
where $m$ is a projection of $j$ on the quantization axes. Let us, for
simplicity, to mark the total set of quantum numbers with $c$. Then the
basis states can be written as follows
\begin{equation} \label{2}
|c \rangle =\left( |nL,s_e(F)s_{\bar{p}}:J\rangle \otimes
Y_l(\mathbf{n})\right)_{jm},
\end{equation}
where  $\mathbf{n} = \mathbf{R}/R$ and $\mathbf{R}$ is a vector
between centers of mass of the colliding subsystems. Total wave
function of the $(\bar{p}\mathrm{He}^+) - \mathrm{He}$ system is
expanded as a series in the basis states $|c\rangle$ with the
coefficients $ \chi_c(R)/R$ depending on the interatomic distance
$R$ and satisfying to the system of coupled-channels equations
\begin{equation}   \label{3}
\left[\frac{d^2}{dR^2} + k_c^2 -\frac{l_c(l_c+1)}{R^2}
\right]\chi_c(R)=2M \sum V_{cc'}(R)\chi_{c'}(R)\,
\end{equation}
where $M$ is a reduced mass of colliding atoms,
$k_c^2=2M[E-(\epsilon_c-\epsilon_i)]$, $E$ is a relative kinetic
energy in the input channel $i$, $\epsilon_c=\epsilon_{FJ}$ is the
energy of the split level,  and $V_{cc'}(R)$ is the matrix element
of the interaction between channel states $c$ and $c'$.

The central part of the interaction between the neutral exotic
atom $(\bar{p}He^+)$ and He atom contains a Van der Waals
attraction ($\sim 1/R^6$) at a large distance and a strong
repulsion at short distances \cite{5,9,10}. Due to non-zero
orbital angular momentum $L$ of the antiprotonic helium, the total
interaction could also include a tensor term that will couple the
states with different projections $\Lambda$ of the angular
momentum $L$. In order to infer a form of the tensor interaction,
let us consider a long-range dipole-dipole interaction between two
atoms
\begin{equation} \label{4}
\hat{V}_1=[(\mathbf{D} \cdot \mathbf{d}) -3 (\mathbf{D} \cdot
\mathbf{n}) (\mathbf{d}  \cdot  \mathbf{n})]/R^3,
\end{equation}
where $\mathbf{D}$ and $\mathbf{d}$ are dipole operators of the
antiprotonic and ordinary atoms, respectively. Matrix elements of
the interaction \eqref{4} between ground states of He atom
disappear, however in the second-order perturbation theory it
generates scalar and tensor terms of an effective Van der Waals
interaction
\begin{equation} \label{5}
\langle nL\Lambda|V(\xi,\mathbf{R})|nL\Lambda' \rangle =-\frac{1}{R^6}
\left(C_6 \delta_{\Lambda \Lambda'} + G_6 \sum_{\nu} \langle L\Lambda'
2\nu|L\Lambda\rangle C^*_{2\nu}(\mathbf{n}) \right) ,
\end{equation}
where  $ C_{2\nu}(\mathbf{n})  =  Y_{2\nu}(\mathbf{n})\sqrt{4\pi/5}$ ,
\begin{align}
 C_6  & =  \frac{2}{3}  \sum_{m\gamma} \frac{|\langle
m|\mathbf{d}|0\rangle|^2 \,
 |\langle \gamma|\mathbf{D}|nL\Lambda \rangle|^2}{E_m -E_0
+\epsilon_\gamma -\epsilon_{nL}},  \label{6} \\
 G_6  & =\frac{1}{2L+1}\sqrt{\frac{2}{3}} \sum_{\lambda\mu\nu\Lambda\Lambda' m\gamma}
 \langle 1\lambda 1\mu |2\nu \rangle
 \langle L\Lambda' 2\nu|L\Lambda \rangle
\frac{|\langle m|\mathbf{d}|0\rangle |^2
 \langle nL\Lambda|D_\lambda|\gamma \rangle
 \langle \gamma|D_\mu|nL\Lambda' \rangle}
{E_m -E_0 +\epsilon_\gamma -\epsilon_{nL}}.  \label{7}
\end{align}
The indexes $nL\Lambda,\gamma$ and $0,m$ refer to the states of
exotic and ordinary atoms, respectively. At large $n$ the density
of the levels is rather large, therefore the main contributions
into the sums over $\gamma$ in \eqref{6} and \eqref{7} arise from
the states with $|\epsilon_\gamma -\epsilon_{nL}| \ll |E_m-E_0|$.
It allows to estimate the scalar and tensor Van der Waals
constants in the closure approximation as
 \begin{align}
 C_6 & =\alpha \langle nL|\mathbf{D}^2|nL \rangle , \label{8} \\
 G_6 & =\alpha \langle nL||\mathbf{D}^2 C_2(\mathbf{D}/D)|| nL \rangle
/\sqrt{2L+1}, \label{9}
 \end{align}
 where $\alpha = 1.383$ a.u. is a static polarizability of He atom.

The simple radial dependence \eqref{5} of the interaction  is valid
only  at long distances. For arbitrary distance we suppose
  \begin{equation} \label{10}
\langle nL\Lambda|V(\mathbf{R},\xi)|nL\Lambda'\rangle= V_0(R)
\delta_{\Lambda\Lambda'} + V_2(R) \sum_{\nu} \langle L\Lambda'
2\nu|L\Lambda\rangle C^*_{2\nu}(\mathbf{n}),
  \end{equation}
where functions $V_0(R)$ and $V_2(R)$ could depend on the quantum
numbers $n,L$. Radial dependence of the scalar term was taken in
the form \cite{4}
 \begin{equation} \label{11}
V_0(R)= -C_6\cdot f(R), \qquad  f(R)= (R^2-r_c^2)/(R^2+r_0^2)^4 ,
 \end{equation}
that has a repulsion at $R < r_c$, Van der Waals minimum at
$R^2_{\min}=(r_0^2+4r_c^2)/3$ and the correct long-range
asymptotic $V_0(R) \rightarrow -C_6/R^6$ at $R\rightarrow\infty$.
Radial dependence of the tensor term at the large distance is
similar, whereas at small distance it has to be $V_2(R)\sim R^2$.
To satisfy these limits we suppose
\begin{equation} \label{12}
V_2(R) = -G_6\cdot f(R)[1-\exp(-\eta R^2)] ,
\end{equation}
 with a large enough value of the parameter $\eta$ ($\sim 10$ a.u.)
in order to avoid its influence at $R\gtrsim 1$ a.u.

\begin{figure}[thb]
      \centering
\includegraphics[width=0.6\textwidth]{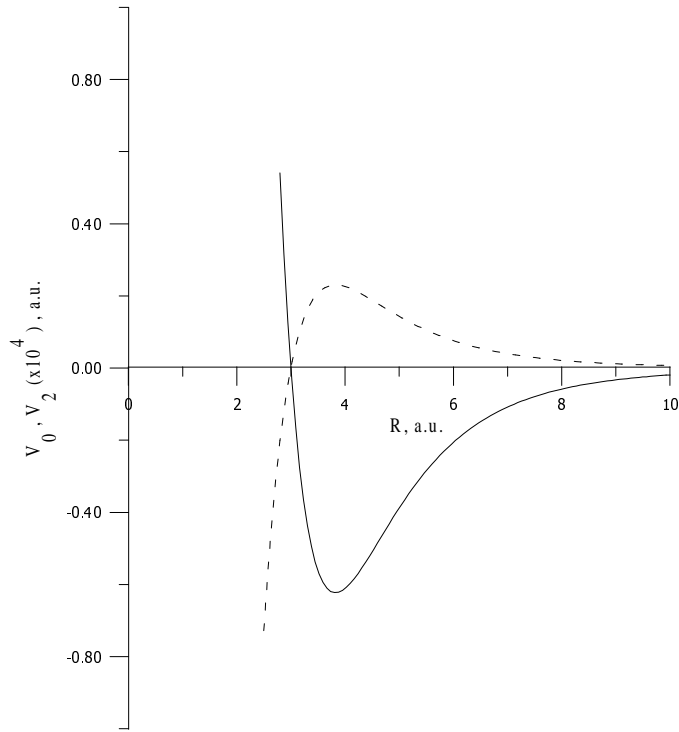}
\vspace*{-3mm}
 \caption{Radial dependence of the scalar $V_0(R)$ (solid line)
and tensor $V_2(R)$ (dashed line) terms of the potential.}
\label{fig:pot}
\end{figure}

For the calculations we use two sets of the parameters. The set A
is based on the fitting  \cite{4} of the density shifts of
E1-transitions $(nL \rightarrow n'L'=L\pm 1)$, $C_6=2.82$, $r_c=
3.0$, $r_0=2.8$ (all values in atomic units). The second set (B)
is estimated using the data of \emph{ab initio} calculations of
the potential  energy surface \cite{5,9}. The repulsion radius,
position and depth of the Van der Waals minimum of the potential
$V_0(R)$ from Fig. 3 in \cite{5} were used to obtain the values
$C_6=3.35$, $r_c= 4.75$, $r_0=0.707$ (a.u.) for our form of the
potential. For the both sets we adopt $G_6/C_6=-0.37$ estimated by
means of a single-particle model of $(\bar{p}\mathrm{He}^+)$ with
effective charges. A dependence of the parameters on $n,L$ is
rather weak and does not matter for our aims. On other hand, two
sets of the parameters differ markedly, that allows to reveal
general properties of characteristics to be considered. General
dependencies of the scalar $V_0$ and tensor $V_2$ terms of the
potential \eqref{10}  on $R$ are shown on Fig. \ref{fig:pot}.

The interaction \eqref{10} does not depend explicitly on the spin
variables, therefore it can change the values of $F$ and $J$ only
due to the vector coupling of the spins with the orbital angular
momentum $L$. The matrix of the potential in the right part of
equation \eqref{3} is reduced to the $3j$ and $6j$-symbols by the
usual Racah algebra:
\begin{equation} \label{13}
\begin{split}
 \langle nLFJ,l:j|V(\mathbf{R},\xi)|nL F'J',l':j \rangle =
  V_0(R) \delta_{FF'}\delta_{JJ'}\delta_{ll'} +
  V_2(R) (-1)^{j+F+F'+L+1} \cdot \\ \cdot  \hat{L} \hat{F} \hat{F'} \hat{J}
\hat{J'} \hat{l} \hat{l'}  
 \left(\begin{matrix} l'& 2 & l \\ 0 &
0 & 0 \end{matrix} \right)
   \left\{\begin{matrix}J & l & j \\ l' & J' & 2 \end{matrix}\right\}
\left\{\begin{matrix}F & J & 1/2 \\ J' & F' & 2 \end{matrix}\right\}
\left\{\begin{matrix}L & F & 1/2 \\ F' & L' & 2 \end{matrix}\right\} ,
\end{split}
\end{equation}
where $\hat{a}\equiv \sqrt{2a+1}$.

With these potentials, we solve coupled channels equations
including 4 HFS sublevels at fixed $nL$ and relative angular
momenta up to $l=5$ at the kinetic energy $E\leq 25$ K and obtain
$S$-matrix in the representation of quantum numbers $F,J,l,j$.
Elastic and inelastic cross sections and  rates of the collisional
transitions between HFS states were calculated with the formulae:
\begin{align}
\sigma(FJ\rightarrow F'J')&=\frac{\pi}{k^2} \sum_{jll'}
\frac{2j+1}{2J+1}\left|\delta_{ll'}\delta_{JJ'}\delta_{FF'} -
 \langle FJl|S^j|F'J'l'\rangle \right|^2, \label{14}       \\
\lambda(FJ\rightarrow F'J') &= N \langle \langle v \sigma(FJ\rightarrow
F'J')\rangle \rangle,           \label{15}
\end{align}
where N is an atomic density of the medium, and the double angular
brackets in the right side of \eqref{15} stand for averaging over the
thermal velocity distribution.

Collisional shifts and broadenings of the microwave M1 transitions
$(FJ \rightarrow F'J')$ can be considered using a general theory
of similar effects in atoms. For non-overlapping levels, Eq.
(57.96) from \cite{11} is relevant to our problem, and, with our
notations of the quantum numbers, becomes
\begin{multline} \label{16}
\gamma + \mathrm{i}\Delta = N\pi \sum_{ll'j_1j_2}
(2j_1+1)(2j_2+1)(-1)^{l+l'}
 \left\{\begin{matrix} j_1 & j_2 & 1 \\ J_2 & J_1 & l \end{matrix}\right\}
 \left\{\begin{matrix} j_1 & j_2 & 1 \\ J_2 & J_1 & l'\end{matrix}\right\}
  \\
 \cdot \langle \langle
vk^{-2} [\delta_{ll'} -  \langle nLF_1J_1 l'|S^{j_1}_I|nLF_1J_1 l
\rangle
  \langle nLF_2J_2 l'|S^{j_2}_{II}|nLF_2J_2 l\rangle ^*]  \rangle  \rangle,
\end{multline}
where S-matrix with the subscript $I$ ($II$) corresponds to the
collisions before (after) the radiative M1 transition $F_1 J_1
\rightarrow F_2 J_2$.

It should be noted that the averaging over the thermal velocity
distribution in Eq. \eqref{16} has to be done in such way that $S_I$
and $S_{II}$ are taken at equal kinetic energies in the corresponding
channels, i.e., at the total energies differed by the HFS splitting,
because radiative transitions $(I\rightarrow II)$ don't change kinetic
energy.  In the computational aspect, it means that for each kinetic
energy $E$ in the channel $F_1,J_1$ we have to solve coupled-channel
equations \eqref{3} four times, at the total energies
$E+\epsilon(F_1J_1) - \epsilon(F_2J_2)$ for all four states $F_2 J_2$.
In spite of the smallness of the splitting, it is comparable with the
thermal energy at low temperature. Moreover,  a presence of resonance
scattering in the considered energy region (see next Section) can make
the effect of the displacement of the total energies in the channels
$I$ and $II$ to be drastically important for the calculations of the
shift and broadening \eqref{16}.
         \section{Results and discussions}
\begin{figure}[thb]
\centering
 \vspace*{-2mm}
\includegraphics[width=0.32\textwidth]{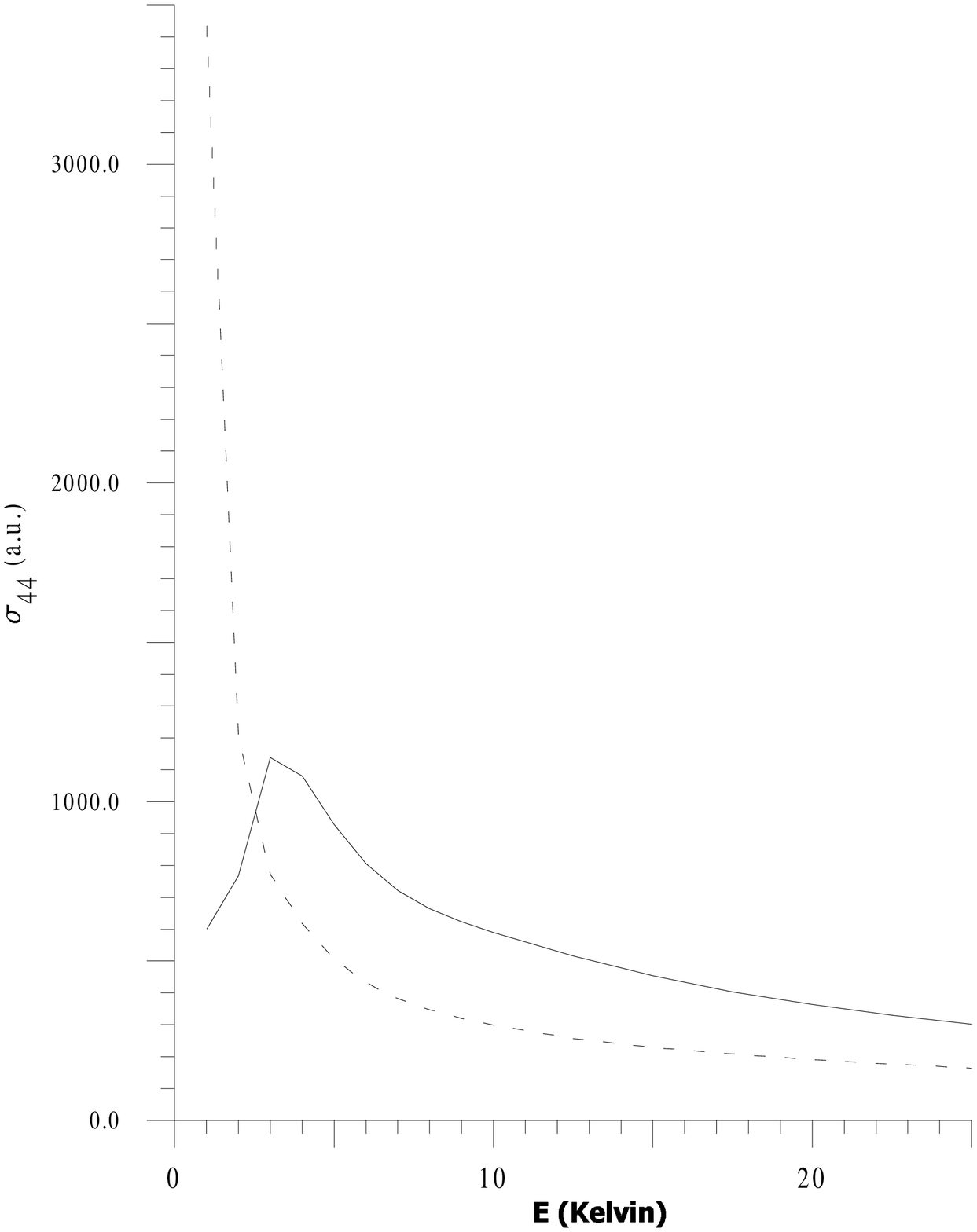}
 \put(-60,150){(a)}
\includegraphics[width=0.32\textwidth]{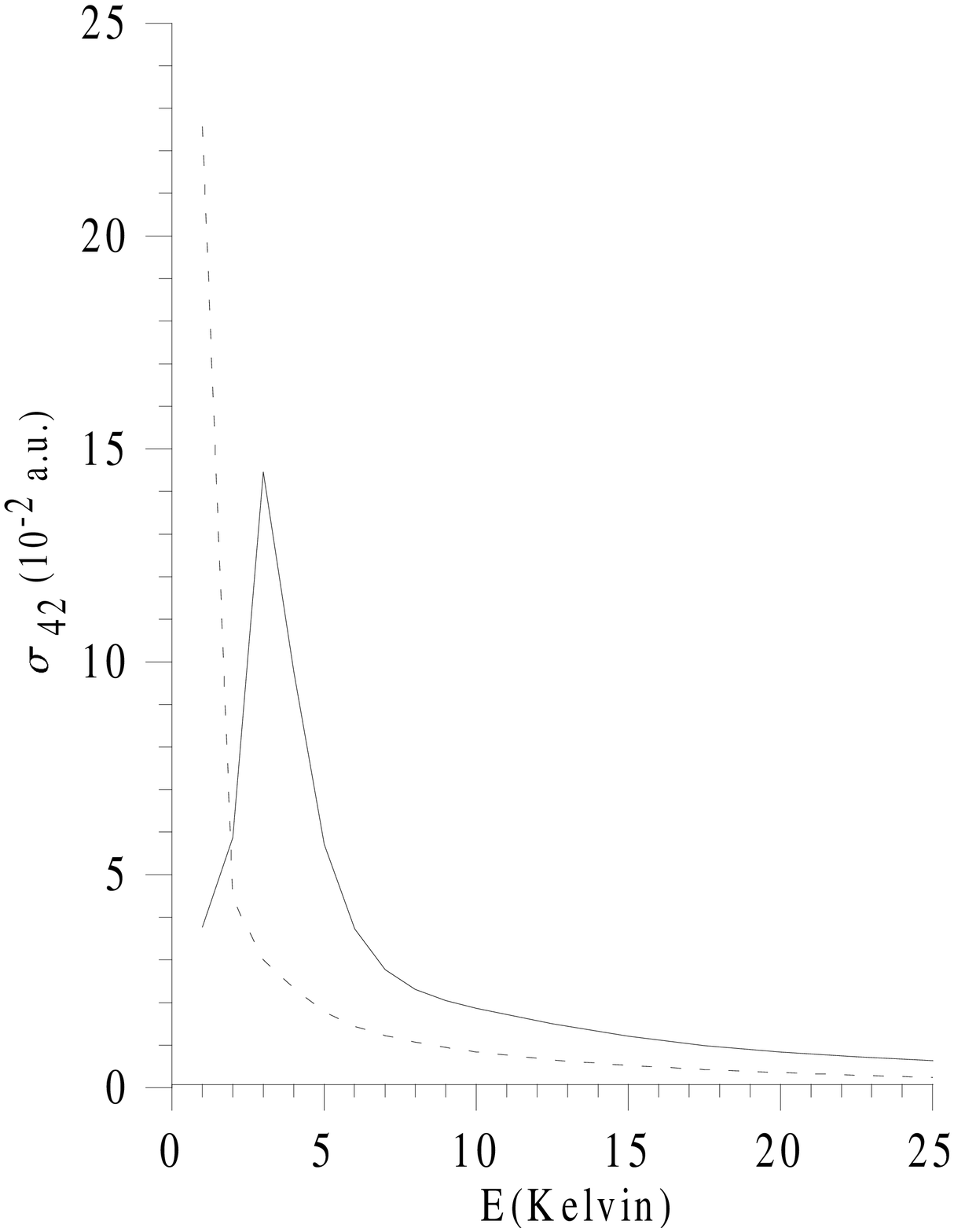}
 \put(-60,150){(b)}
\includegraphics[width=0.32\textwidth]{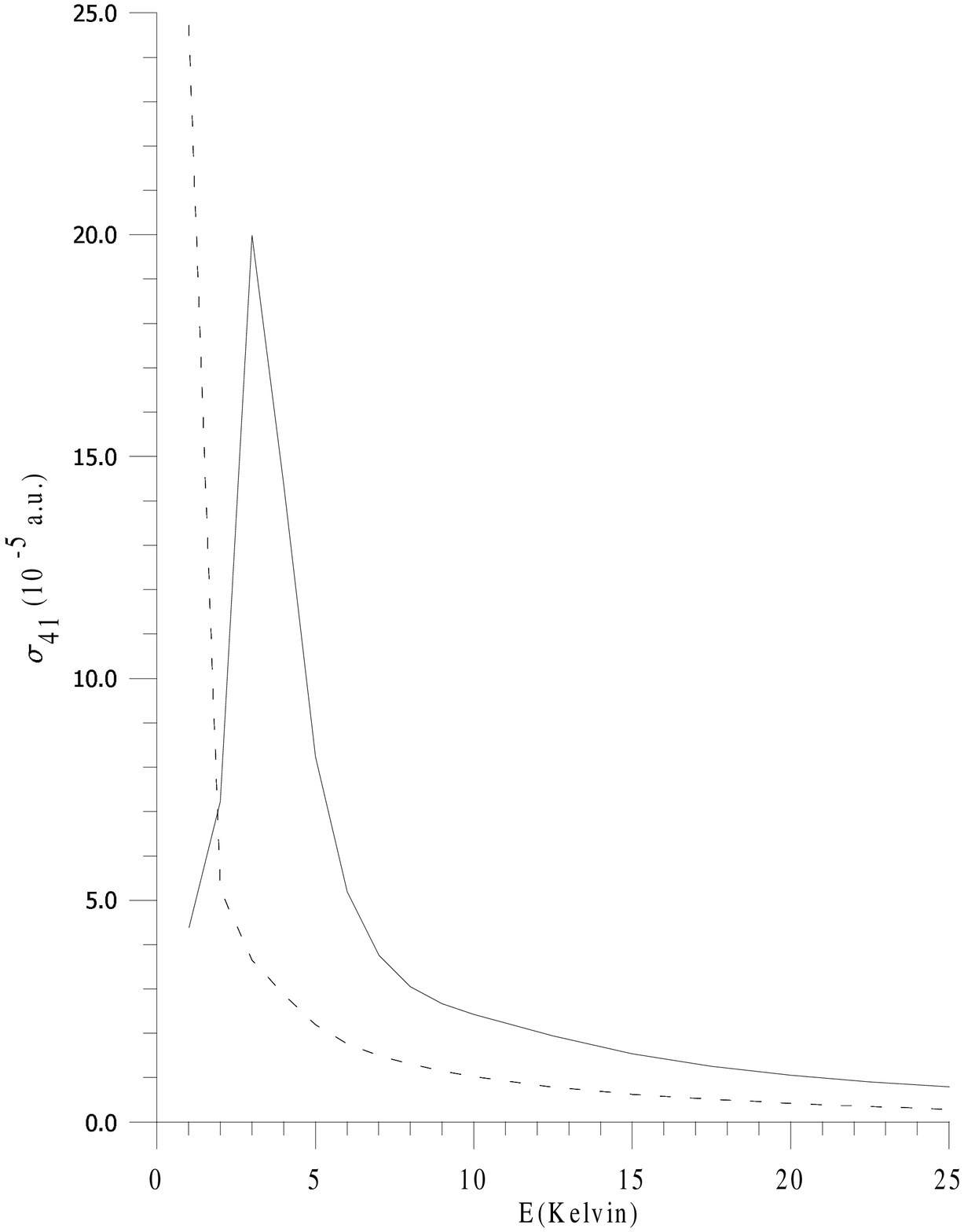}
 \put(-60,150){(c)} 
 \vspace*{-5mm}
 \caption{Cross sections of $(\bar{p}\mathrm{He}^+)_{nL} -\mathrm{He}$
collisions  \emph{vs.} c.m. kinetic energy: (a) elastic cross section
$\sigma_{44}$ for the initial HFS state
$|4\rangle=|F=L-1/2,J=F+1/2\rangle$; (b) inelastic (de-excitation)
cross section  $\sigma_{42}$ for the 'electron spin-flip' transition
$4\rightarrow 2$; (c) inelastic (de-excitation) cross section
$\sigma_{41}$ for the 'electron and antiproton spin-flip' transition
$4\rightarrow 1$. Solid  and dashed lines refer to the sets A and B of
the parameters of the potential.} \label{fig3:sigma}
\end{figure}

\begin{figure}[thb]
\centering
 \vspace*{-5mm}
\includegraphics[width=0.32\textwidth]{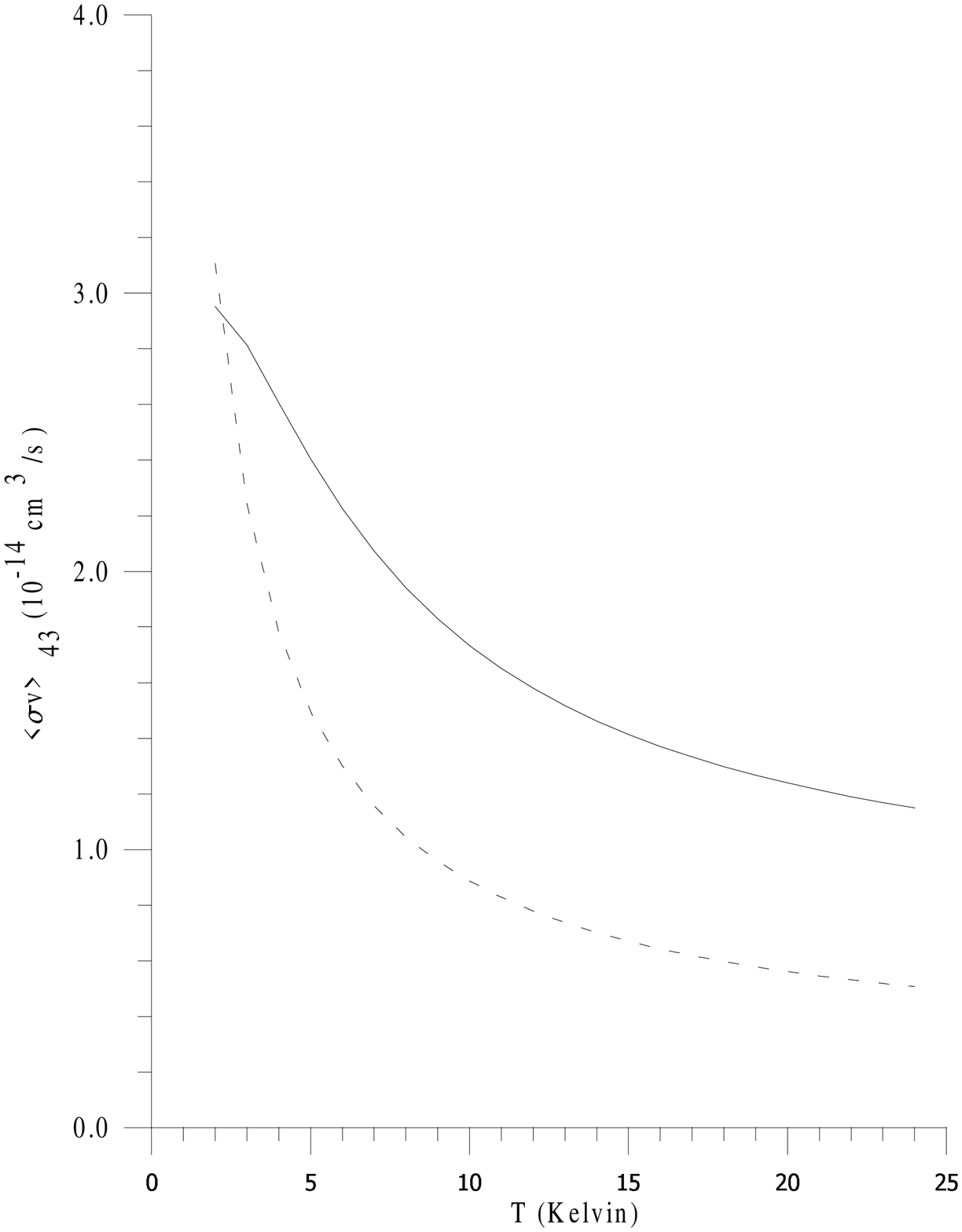}
 \put(-60,150){(a)}
\includegraphics[width=0.32\textwidth]{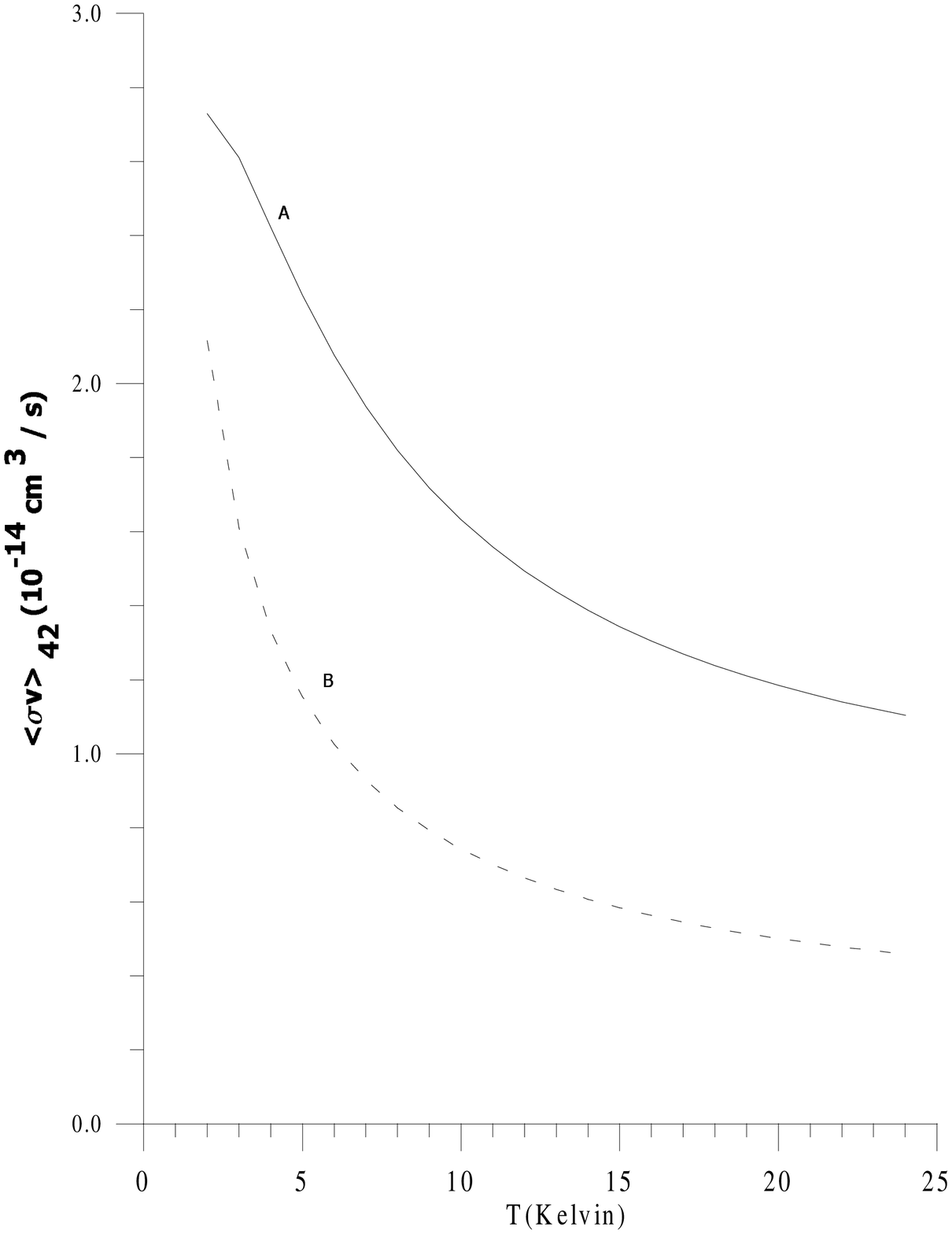}
  \put(-60,150){(b)}
\includegraphics[width=0.32\textwidth]{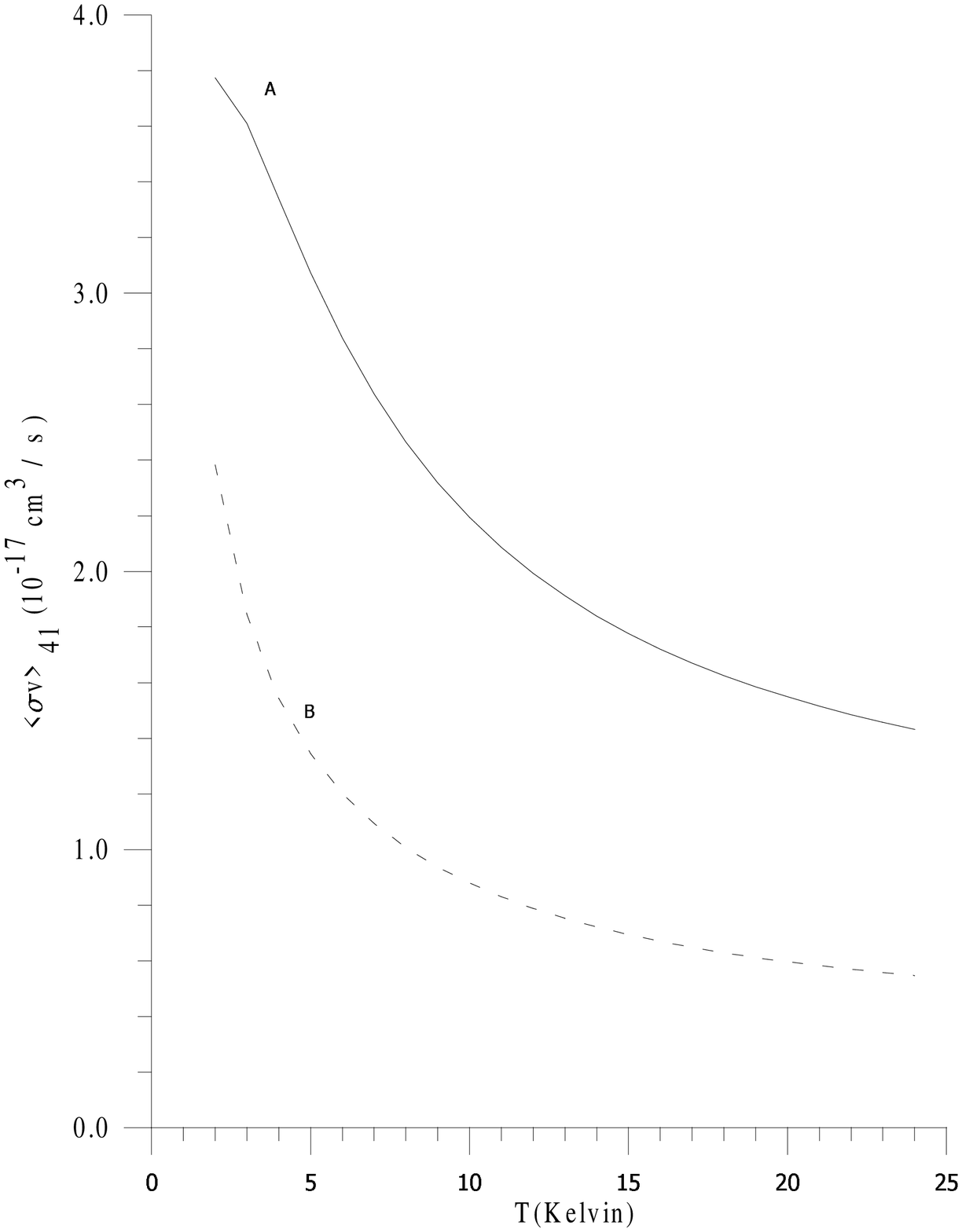}
  \put(-60,150){(c)}
 \vspace*{-5mm}
\caption{Temperature dependence of per-atom collisional transition
rates $\lambda(i\rightarrow j)/N=\langle \langle\sigma(i\rightarrow j)
v\rangle \rangle$; the figures (a), (b) and (c) are for the transitions
$4 \rightarrow 3$, $4 \rightarrow 2$ and $4 \rightarrow 1$,
respectively. The solid and dashed lines are obtained with the sets A
and B of the parameters of the potential \eqref{11} - \eqref{12}.}
\label{fig4:rate}
\end{figure}

\begin{figure}[thb]
\centering
 \vspace*{-8mm}
\includegraphics[width=0.48\textwidth]{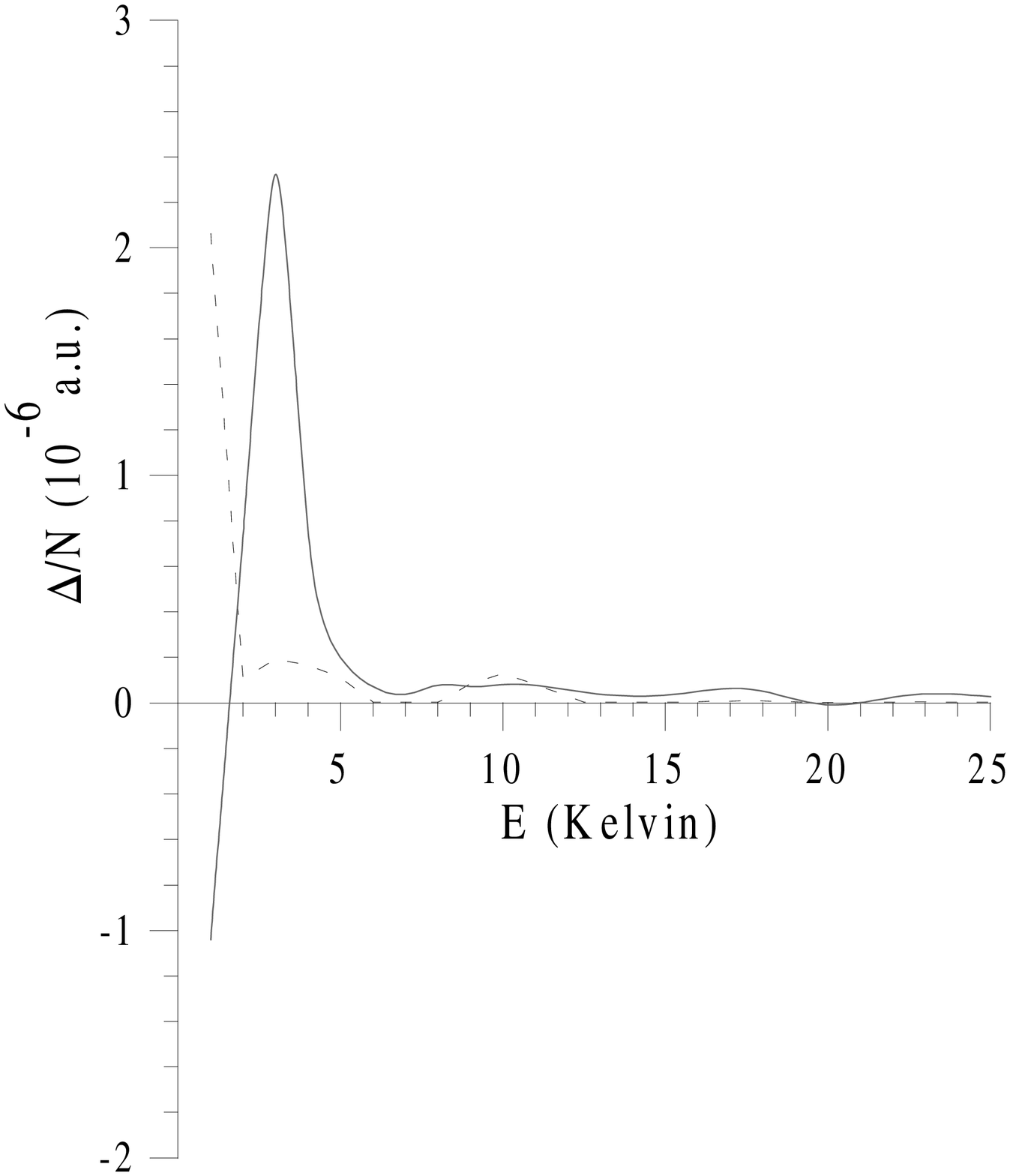}
 \put(-70,200){(a)}
\includegraphics[width=0.48\textwidth]{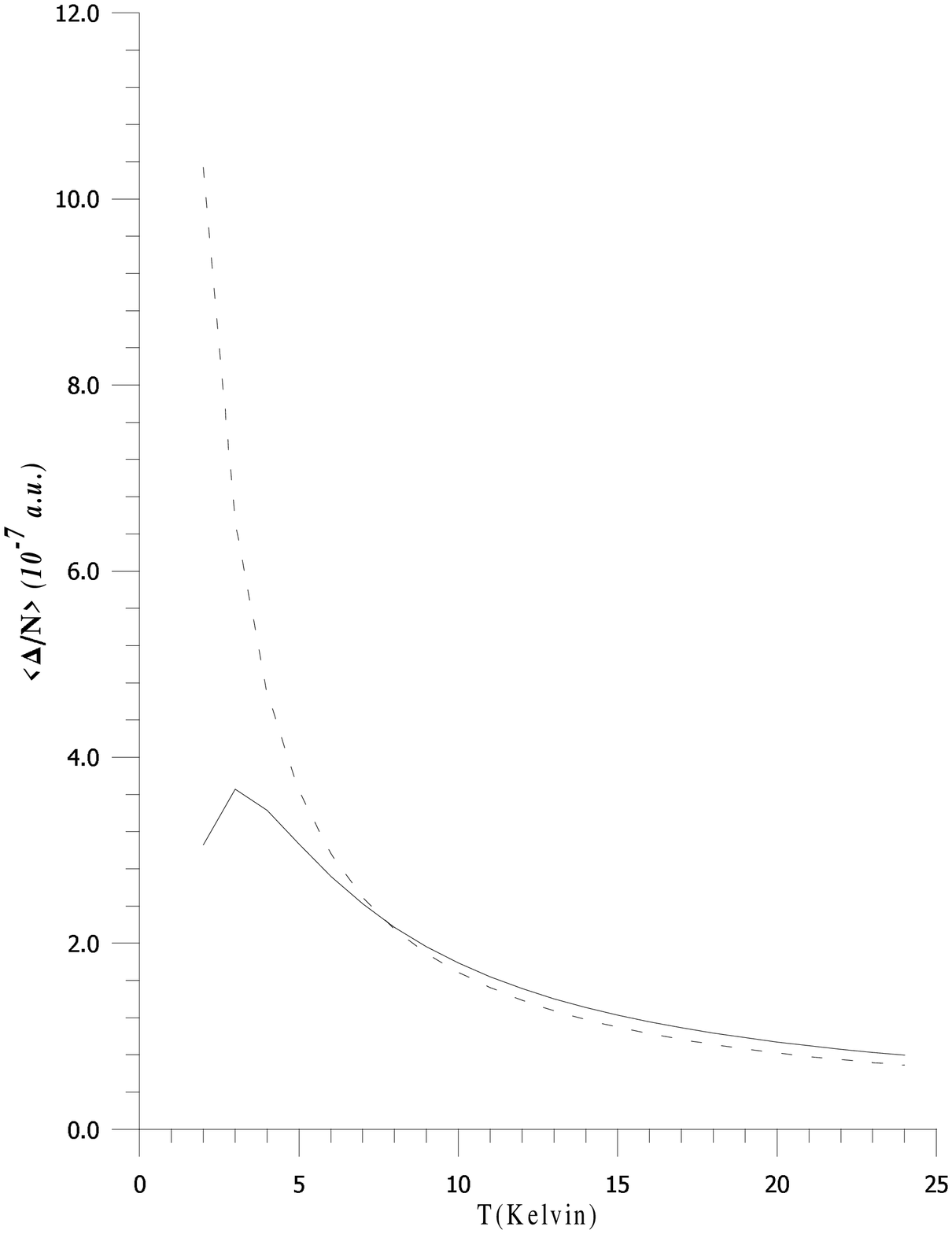}
 \put(-70,200){(b)}
\vspace*{-5mm}
 \caption{Per-atom density shift $\Delta/N$ of the M1 spectral
line $4 \rightarrow 2$: (a) dependence on kinetic energy before
averaging over Maxwell distribution, (b) dependence on temperature. The
solid and dashed lines correspond to the sets A and B of the parameters
of the potential. The curves for the shift of $(3\rightarrow 1)$
spectral line are indistinguishable from the shown for the $4
\rightarrow 2$ spectral line.} \label{fig5:shift}
\end{figure}

\begin{figure}[thb]
\centering
  \vspace*{-8mm}
\includegraphics[width=0.48\textwidth]{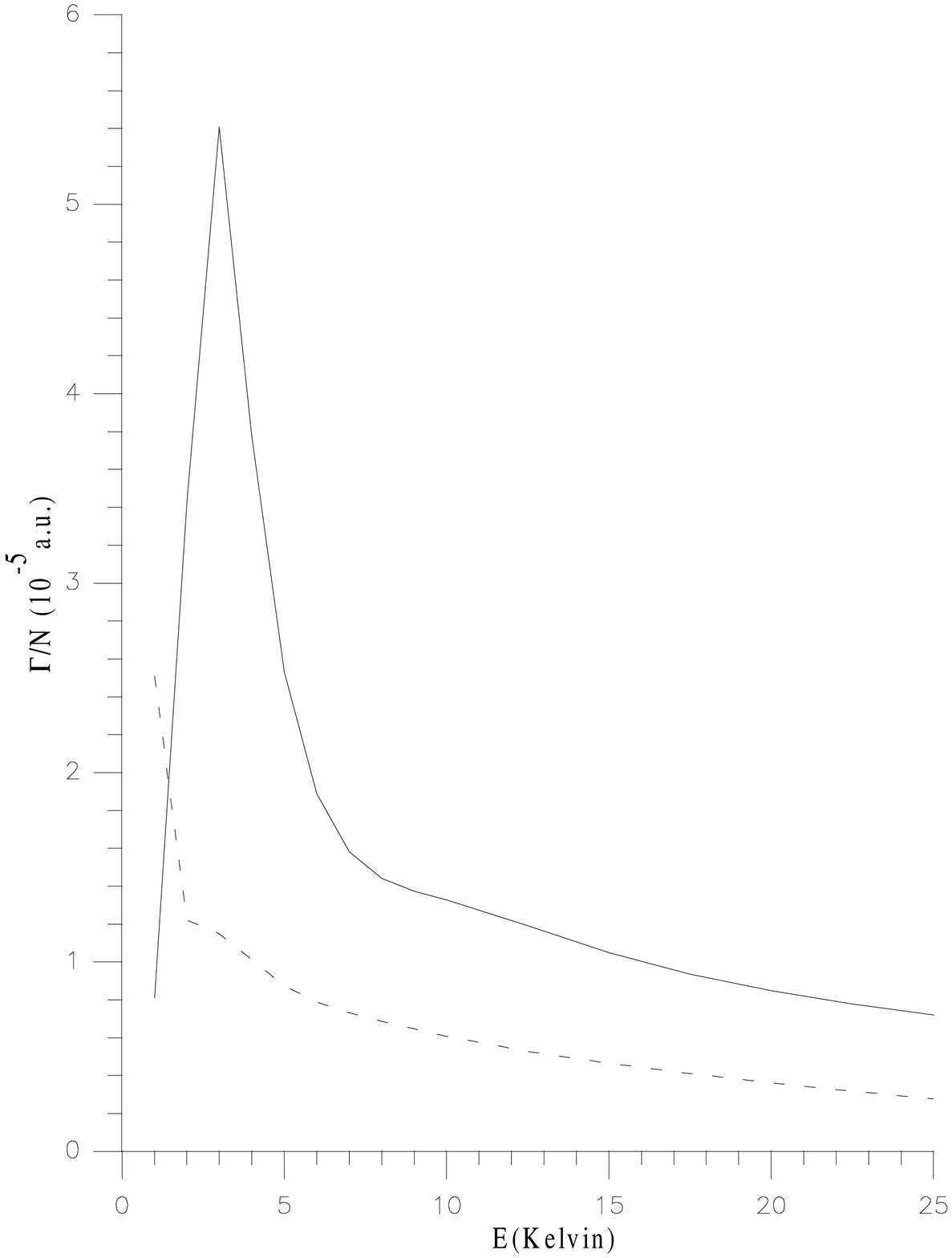}
  \put(-70,200){(a)}
\includegraphics[width=0.48\textwidth]{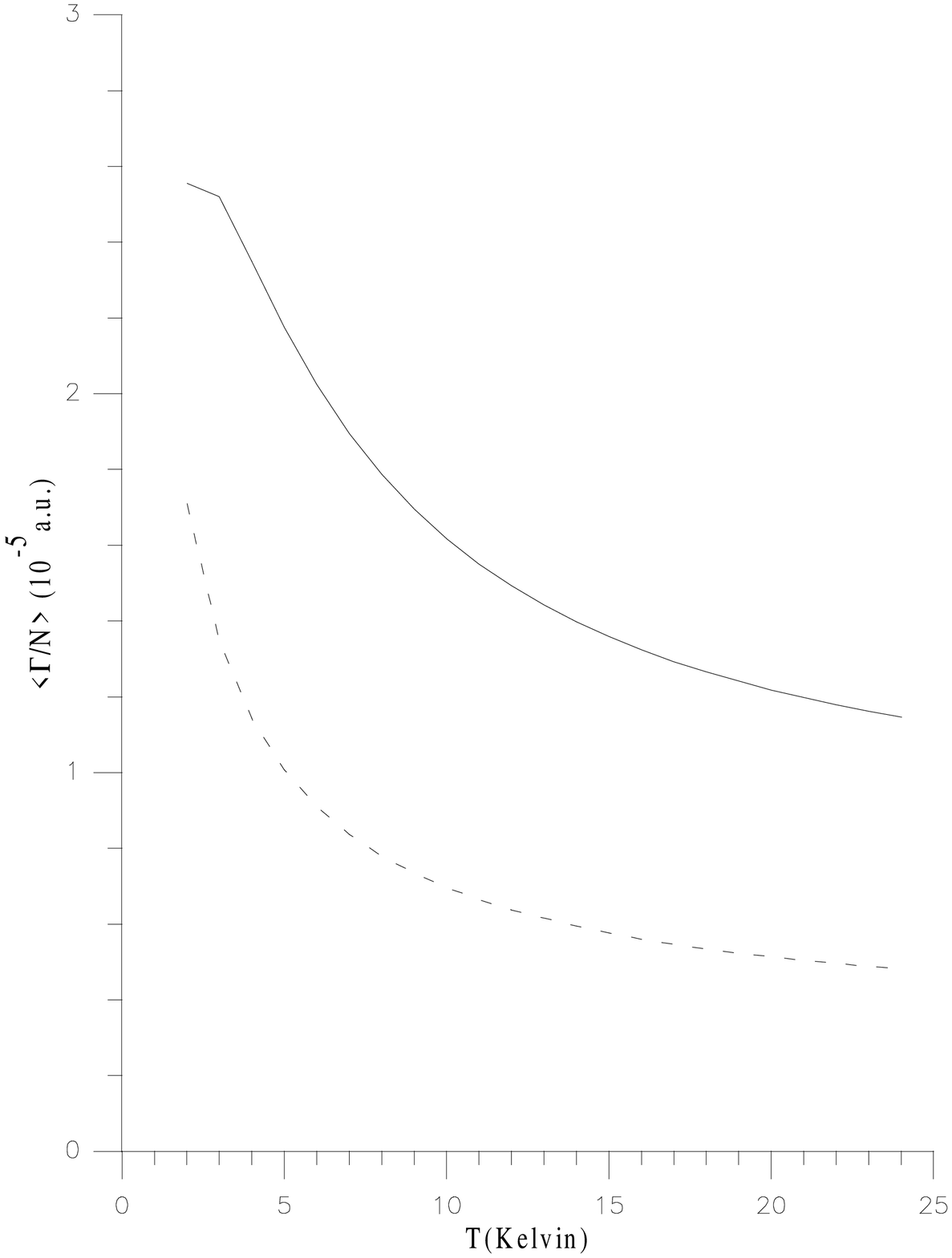}
  \put(-70,200){(b)}
 \vspace*{-5mm}
 \caption{Per-atom density broadening $\Gamma/N$ of the M1 spectral
line $4 \rightarrow 2$: (a) dependence on kinetic energy before
averaging over Maxwell distribution, (b) dependence on temperature. The
solid and dashed lines correspond to the sets A and B of the parameters
of the potential. The curves for the shift of $(3\rightarrow 1)$
spectral line are indistinguishable from the shown for the $4
\rightarrow 2$ spectral line.} \label{fig6:gamma}
\end{figure}

The values of elastic and inelastic cross sections \eqref{14}, rates of
collisional transitions \eqref{15}, shifts and broadenings \eqref{16}
of the M1 spectral lines due to collisions of $(\bar{p}\mathrm{He}^+)$
with $\mathrm{He}$ atoms were calculated in this paper for all possible
pairs of initial and final states $(FJ,F'J')$.  Typical results of the
calculations are shown on Figs. \ref{fig3:sigma} - \ref{fig6:gamma}.

It should be noted that elastic cross sections are practically
coincide for all four states $|FJ\rangle$ at the same kinetic
energy in the corresponding channel, so they are indistinguishable
on the figures. Much of the same is true for the pair of inelastic
cross sections for the transitions with the spin-flip similarity,
e.g., $\sigma(4\rightarrow 2)\simeq \sigma(3\rightarrow 1)$,
$\sigma(4\rightarrow 3)\simeq \sigma(2\rightarrow 1)$. Therefore
we show in Figs. \ref{fig3:sigma} (a, b, c) elastic and inelastic
cross section for the single initial state $(F=L-1/2,J=F+1/2)$
(number 4 on the energy diagram \ref{fig:HFS}). Solid lines show
the cross sections obtained with the set A of the parameters of
the interaction potential, dashed lines show the similar
cross-sections obtained with the set B of the parameters.
Considering these results as well as the results for another
initial states, we can see that elastic and inelastic cross
sections have a similar energy dependence, but differ in magnitude
by several orders of value. More definitely, inelastic cross
sections corresponding to the single spin-flip (electron spin-flip
$\Delta F =\Delta J =\pm 1$ or antiproton spin-flip $\Delta F=0,
\Delta J =\pm 1$) are less than elastic cross sections by four
orders of value, whereas inelastic cross sections for the double,
electron and antiproton, spin-flip ($\Delta F=\pm 1,\, \Delta J=0$
or $\Delta J=\pm 2$) are less than elastic cross sections by seven
orders of value.

Also, it is seen from Fig. \ref{fig3:sigma} that the cross
sections are essentially dependent on the parameters of the
potential. So, an investigation of the elastic cross section can,
in principle,  shed light on the potential of interaction between
antiprotonic and ordinary atoms. However, the most interesting
feature of the cross sections on this figure is a pronounced
maximum for all cross sections at a low energy (about $3$ K for
the set A of the parameters). This maximum is presented in all
channels of the elastic and inelastic scattering and therefore
should be due to the existence of resonances (quasistationary
levels). Analysis of the results reveals the resonances in $S$,
$P$ and $D$-partial waves at the nearly equal energy ($\simeq 3$
K). Appearance of these quasistationary states is due to the
radial dependence of potential (see Fig. \ref{fig:pot}), which has
a wide region of the attraction bounded by a repulsive barrier at
$R=r_c$ and by a centrifugal barrier at large $R$. This part of
the potential in principle could ensues the existence of
quasistationary levels. A possible existence of the resonances as
well as of bound states in the system in the
$(\bar{p}\mathrm{He}^+)_{nL} - \mathrm{He}$ was discussed for the
first time in the paper \cite{4} with account for the scalar term
$V_0(R)$ of the potential. Recently this prediction was confirmed
in the calculations with an averaged \emph{ab initio} potential
\cite{12}. It is of the interest that for the set B of the
parameters of the potential the resonance is shifted to lesser
energies and appears itself as the strong growth of the
cross-section at lesser energies.

The rates of transitions between HFS levels induced by collisions
of the  $(\bar{p}\mathrm{He}^+)_{nLFJ}$ with $\mathrm{He}$ atoms
are connected with the cross sections by Eq. \eqref{15}. Our
results for the values $\lambda(i\rightarrow j)/N$ depending on
temperature are shown on Fig. \ref{fig4:rate} for the transitions
$4\rightarrow 3$, $4\rightarrow 2$ and $4\rightarrow 1$. Solid and
dashed lines are obtained with the sets A and B of the potential
parameters, respectively. The rates of the single spin-flip
transitions $4\rightarrow 3$ and $4\rightarrow 2$ are of the same
order of value and rather large ($\sim N\cdot 10^{-14}
\mathrm{cm}^3/\mathrm{s}$), whereas the rates of double spin-flip
transitions are suppressed by three orders of value. For an
interpretation of the experiment \cite{6} the most interesting
collisional transitions are those that change relative populations
of upper (4 and 3) and lower (2 and 1) groups of the levels. We
have noted in Introduction that it follows from the experiment
that a relaxation time of the populations has to be of order or
greater than a time gate between two laser pulses, i.e., 140 ns.
This value should be compared with inverse rates of the
transitions $4\rightarrow 2$, $4\rightarrow 1$,  $3\rightarrow 2$
and $3\rightarrow 1$. According to our calculations,
$\lambda(3\rightarrow 1)\simeq \lambda(4\rightarrow 2)$, whereas
the rates of two other transitions are negligible. At the
experimental conditions ($T=6$ K, $N=3\times 10^{20}$  cm$^{-3}$)
we obtain a collisional relaxation time
$\tau_c=1/\lambda(4\rightarrow 2)= 160.5$ ns for the set A of the
parameters, and $\tau_c= 325$ ns for the set B. The both values
are greater than the mentioned time gate and, so, are compatible
with the experiment.

It is seen from  Fig. \ref{fig4:rate} that the resonances existing
in the cross sections are not presented here in the form of
maximum, but, due to averaging over the thermal motion, manifest
itself as a reduction of the rates with increasing of the
temperature: all considered rates $\lambda(FJ\rightarrow F'J')$
are reduced at least by factor 2 for the both sets (A and B) of
the parameters when the temperature raises from $T=3$ K to 25 K.
In other words, the relaxation time of the populations of the HFS
levels has to rise in the considered temperature region by factor
about 2. This result may be especially interesting for the further
experiments with HFS of the antiprotonic helium by the triple
laser-microwave-laser resonance method, because a such temperature
dependence of the relaxation time allows to find more convenient
conditions for the time gate between two laser pulses. Of course,
we can not extrapolate the revealed T-dependence to higher
temperature without further calculations. We took into account in
these calculations the partial waves up to $l=5$ that is enough in
the considered region ($T<25$ K). However, at higher temperature
we can expect, from general point of view, that the main
contribution to the rates far from the resonance region will give
higher partial waves, and a number of essential partial waves will
increase with the mean energy, therefore the rates could increase
with temperature.

Per-atom density shift $\Delta/N$ and broadening $\Gamma/N$ of the
frequency of M1 spectral line $4 \rightarrow 2$ are shown on the Figs.
\ref{fig5:shift} and \ref{fig6:gamma}, respectively. Left parts (a) of
the figures display  dependencies of the shift and broadening on
kinetic energy (before averaging over Maxwell distribution), and right
parts (b) show dependencies on temperature (with averaging over the
Maxwell distribution). The solid and dashed lines correspond to the
sets A and B of the parameters of the potential. The same curves could
be referred to the $(3\rightarrow 1)$ spectral line, because the
results are indistinguishable in the scale of the figures.

A dependence of the shift of spectral lines on kinetic energy of the
collision comes about as a result of interference of different partial
waves, including resonant and non-resonant ones, as can be seen from
the imaginary part of Eq. \eqref{16}. Thus, it is understood some
oscillations in the curves for the shift that one can see on the Fig.
\ref{fig5:shift}a. After averaging over Maxwell distribution on
velocity, the oscillations disappear, and temperature dependence of the
shift shows more smooth curves. Nevertheless, it is of some interest
that the overall form of curves is defined by existence of resonance
region at energy of several K.

On other side, broadening of the spectral lines reveals a resonance
energy dependence (see Fig. \ref{fig6:gamma}a) similar to that one of
the cross sections, because the real part of Eq. \eqref{16} contains
incoherent contributions of different partial waves.  After averaging
over thermal motion we obtain the temperature dependence of the
broadening (Fig. \ref{fig6:gamma}b), which is rather flat and decreases
with temperature, as it recedes from the resonance region, similarly to
the relaxation rates discussed above.

Considering shift and broadening at the experimental conditions
($T=6$ K, $N=3\times 10^{20}$  cm$^{-3}$), we have found the
frequency shift of M1 spectral lines $\Delta\nu\simeq 79.5$ kHz
for the favored transitions ($\Delta F=\pm 1,\, \Delta J=\pm 1$)
with the set A, and rather close (86.6 kHz) with the set B of the
potential parameters. This value is less than experimental
accuracy ($\simeq$300 kHz) of the measured values
$\nu_{HF}^{\pm}$, therefore the density shift could not be
revealed in the experiment \cite{6}. The calculated total
broadening at the mentioned conditions for the same spectral lines
is $\gamma=5.92$ MHz and  2.66 MHz for the set A and B of the
parameters, respectively. The both values are compatible with the
experimental value of the broadening for the favored spectral
lines ($5.3\pm 0.7$ MHz).

\section{Conclusion}

In this paper we have investigated a set of the effects of
$(\bar{p}\mathrm{He}^+)_{nLFJ} - \mathrm{He}$ collisions on
transitions between hyperfine-structure sublevels in the frame of
coupled channel approach. We have taken into account four
hyperfine-splitting states of the antiprotonic atom with quantum
numbers n, L. With account for six  states of relative orbital
angular momentum, we have got twenty four coupled channels. A
coupling between different channels was calculated in the model of
Van der Waals forces with parameters obtained in two different
approximations. Within this approach we have calculated the
elastic and inelastic cross sections for all four states of the
antiprotonic atom, as well as statistically averaged rates of
transitions between HFS states, the energy and temperature
dependencies of the shifts and broadenings of the M1 spectral
lines for different transitions. We consider that the main
interest of our paper is in the consistent calculation of the
whole set of collision characteristics of antiprotonic atoms with
helium atoms of medium. For the cross sections in the elastic and
inelastic channels we have observed the resonance region that is
due to existence of quasistationary states in the chosen
potential. Effects of the resonances should disappear after
averaging over Maxwell distribution. This appears in the curves
for the rates of the elastic and inelastic scattering. In this
case one can guess about the existence of resonance region only by
the decrease of rates at energies above 5 K. We have calculated
the shifts and broadenings of M1 spectral lines as functions of
energy and temperature. The shifts and broadenings is
characterized with the strong maximum that is due to existence of
resonances at the energies of the order 3-5 K. After averaging
over the Maxwell distribution these maximum disappear that is
quite naturally.

At the density and temperature corresponding to experimental
conditions \cite{6} ($N=3\times 10^{20}$ cm$^{-3}$ and $T=6$ K) we
have obtained the relaxation times $\tau(FJ\rightarrow F'J')\geq
160$ ns, the frequency shifts of M1 spectral lines
$\Delta\nu\simeq 80$ kHz for the favored transitions ($\Delta
F=\pm 1,\, \Delta J=\pm 1$) and frequency broadening of the M1
spectral lines $\gamma \lesssim 5.9$ MHz. These results are
compatible with the experimental data obtained by a
laser-microwave-laser resonance method. With the temperature
rising up to 25 K the rate of relaxation $\lambda=1/\tau$ as well
as shift and broadening of the M1 microwave lines are lowered by a
factor $1.5 \div 2 $ that may be useful a choice of conditions in
further experiments.
\section*{Acknowledgements}
This investigation was supported by Russian Foundation for Basic
Research by the grant 03-02-16616. Authors thank to N.P. Yudin for the
participation in the first stage of the work and for numerous
discussions. One of the authors (G.K.) thanks to T. Yamazaki, R. Hayano
and E. Widmann for attracting our interest to the considered problem,
fruitful discussions and eliminating information on the experiment.

\end{document}